\documentclass[prd, 11pt,a4paper,twocolumn,groupedaddress,preprintnumbers,nofootinbib,floatfix]{revtex4-1}
\usepackage[top=3cm, bottom=2.1cm, left=2cm, right=2cm]{geometry}
\pdfoutput=1

\usepackage{comment}

\usepackage{graphicx}
\usepackage{url}
\usepackage[bookmarks, pagebackref=false]{hyperref}
\usepackage[usenames,dvipsnames]{xcolor}
\usepackage{dcolumn}% Align table columns on decimal point
\usepackage{bm}% bold math
\usepackage{bbm}
\usepackage{amsmath,amssymb,amsfonts}
\usepackage{color}
\usepackage{hyperref}
\usepackage[Symbolsmallscale]{upgreek}
\usepackage{dsfont}
\usepackage[vcentermath]{youngtab}
\usepackage[all]{xy}
\usepackage{pstricks}
\usepackage{dsfont}%
\usepackage{mathtools}
\usepackage{placeins}
\usepackage{enumerate}
\usepackage{cases}
\usepackage{placeins}
\usepackage{xspace}
\usepackage{cancel} % to make the barred text notation
\usepackage{slashed}
\usepackage[caption=false, labelformat=simple, listofformat=subsimple, labelfont=default, margin=5pt,
    justification=raggedright]{subfig}
\usepackage{natbib}

\newcommand{\beq}{\begin{eqnarray}}
\newcommand{\eeq}{\end{eqnarray}}

\newcommand{\bmp}{\noindent\begin{minipage}{16cm}}
\newcommand{\emp}{\end{minipage}\vskip 7mm} % 7mm untightened

    \newcommand{\ii}{\mathrm{i}}

    \newcommand{\SU}{\mathrm{SU}} 
    \newcommand{\Sp}{\mathrm{Sp}}
    \newcommand{\SO}{\mathrm{SO}}
    \newcommand{\Tr}{\mathrm{Tr}}
    \newcommand{\EB}{E_\mathrm{B}}
    \newcommand{\EE}{E_{\theta}}

    \newcommand{\bee}{\begin{equation}}
        \newcommand{\eee}{\end{equation}}

\newcommand{\XE}[3]{{#1}^{#2}\cdot {#3}}

\def\lsim{\mathrel{\rlap{\lower4pt\hbox{\hskip1pt$\sim$}}
    \raise1pt\hbox{$<$}}}                % less than or approx. symbol
\def\gsim{\mathrel{\rlap{\lower4pt\hbox{\hskip1pt$\sim$}}
    \raise1pt\hbox{$>$}}}                % greater than or approx. symbol

\newcommand{\bphi}{\bm\varphi}

\begin{document}

\title{Vacuum alignment with(out) elementary scalars}
\author{Tommi {\sc Alanne}}
\email{alanne@cp3.sdu.dk}
\affiliation{{CP}$^{ \bf 3}${-Origins} \& the Danish Institute for Advanced Study {\rm{Danish IAS}},  University of Southern Denmark, Campusvej 55, DK-5230 Odense M, Denmark.}
\author{Helene {\sc Gertov}}
\email{gertov@cp3.sdu.dk} 
\affiliation{{CP}$^{ \bf 3}${-Origins} \& the Danish Institute for Advanced Study {\rm{Danish IAS}},  University of Southern Denmark, Campusvej 55, DK-5230 Odense M, Denmark.}
\author{Aurora {\sc Meroni}}
\email{meroni@cp3.sdu.dk}
\affiliation{{CP}$^{ \bf 3}${-Origins} \& the Danish Institute for Advanced Study {\rm{Danish IAS}},  University of Southern Denmark, Campusvej 55, DK-5230 Odense M, Denmark.}
\author{Francesco {\sc Sannino}}
\email{sannino@cp3.dias.sdu.dk}
\affiliation{{CP}$^{ \bf 3}${-Origins} \& the Danish Institute for Advanced Study {\rm{Danish IAS}},  University of Southern Denmark, Campusvej 55, DK-5230 Odense M, Denmark.}

\begin{abstract}
We systematically elucidate differences and similarities of the vacuum alignment issue in composite and  renormalizable elementary  extensions of the Standard Model featuring a pseudo-Goldstone Higgs. We also provide general conditions for the stability of the vacuum in the elementary framework, thereby extending previous studies of the vacuum alignment. 
  \\
[.1cm]
{\footnotesize  \it Preprint: CP$^3$-Origins-2016-037 DNRF90 }
\end{abstract}
\maketitle
\newpage

%%%%%%%%%%%%%%%%%%%%%%%%%%%%%%%%%%%%%%%%%%%%%%%%%%%%%%%%%%%%%%%%%%%%%%%%%%%
\section{Introduction}
\label{sec:ingredients}
%%%%%%%%%%%%%%%%%%%%%%%%%%%%%%%%%%%%%%%%%%%%%%%%%%%%%%%%%%%%%%%%%%%%%%%%%%%

Theories in which the gauge group is only a part of the original global symmetry can undergo a vacuum (mis)alignment phenomenon via quantum corrections. In two pioneering papers in the context of technicolor, Peskin~\cite{Peskin:1980gc} and Preskill~\cite{Preskill:1980mz}  recognized that quantum corrections stemming from the electroweak (EW) sector can destabilize the original vacuum endangering the EW stability. Later Kaplan and Georgi  \cite{Kaplan:1983fs,Kaplan:1983sm} turned the technicolor misalignment problem into a feature by realising that the Higgs doublet of the Standard Model (SM)  can be identified with a doublet of dynamically generated Goldstone bosons (GBs). The final vacuum of the theory depends heavily on the details of the dynamics  responsible for generating the SM fermion masses which, for purely fermionic composite extensions, is unknown.

 We start with reviewing well-known results in the composite framework,
and then show how the analysis modifies when considering elementary scalars. In the elementary case we start from the general form of the Coleman--Weinberg (CW) effective potential \cite{Coleman:1973jx} and then deduce general conditions that, once satisfied, lead to the proper vacuum alignment.
  
The paper is organized as follows: In Sec. \ref{sec:composite} we review the general framework for composite realizations of the pseudo-Goldstone Higgs scenario and associated vacuum alignment conditions. The elementary counterpart is investigated in Sec.~\ref{sec:elementary}. There we setup the framework and provide the general form of the quantum corrections along with deriving relevant conditions for the vacuum alignment. We then apply our results to relevant phenomenological examples in Sec.~\ref{sec:so5}. We finally conclude in Sec.~\ref{sec:conclusions}.

%%%%%%%%%%%%%%%%%%%%%%%%%%%%%%%%%%%%%%%%%%%%%%%%%%%%%%%%%%%%%%%%%%%%%%%%%%%
\section{Review of the vacuum alignment in composite scenarios}
\label{sec:composite}
%%%%%%%%%%%%%%%%%%%%%%%%%%%%%%%%%%%%%%%%%%%%%%%%%%%%%%%%%%%%%%%%%%%%%%%%%%%

Let us consider underlying composite framework with global chiral-symmetry-breaking pattern $\SO(N)\rightarrow\SO(N-1)$. The corresponding
coset space is parameterized by 
\begin{equation}
    \label{eq:}
    \Sigma=\exp\left(\frac{\ii}{f}\Pi^aX^a\right)E,
\end{equation}
where the $\Pi^a$ are the GBs corresponding to the broken generators $X^a$, $f$ is the `pion decay constant', and $E$ gives the vacuum orientation inside $\SO(N)$.

    %%%%%%%%%%%%%%%%%%%%%%%%%%%%%%
    \subsection{The gauged subsector and fermion embedding}
    %%%%%%%%%%%%%%%%%%%%%%%%%%%%%%
	
	Let $K\leq \SO(N)$ be gauged, and assume  that the vacuum alignment $E_0$ preserves $K$, whereas $\EB$ breaks it maximally
	to $K_{\mathrm{B}}$.  In the following, we investigate the alignment of the vacuum  with respect to the gauged subgroup $K$.   

	To this end, we parameterise  $E$ as: 
	\begin{equation}
	    \label{eq:Etot}
	    E\equiv \EE=\cos\theta E_0+\sin\theta\EB,
	\end{equation}
	where the angle $\theta$ is \emph{a priori} a free parameter in the range $[0, \pi/2]$. 
	
	Let us start by defining the basis with respect to the gauge-breaking vacuum $\EB$. Let $\{\tau^a\}$ be the set of generators of the gauge group $K$, 
	which can be divided 
	into the broken, $\{\chi^a\}$, and unbroken generators, $\{\sigma^a\}$. 
	Since $E_0$ preserves the full gauge group, both $\sigma^a$ and $\chi^a$ leave the vacuum $E_0$ invariant. Furthermore, 
	the unbroken $\sigma^a$ also leaves $\EB$ %, and thus the full vacuum, 
	invariant, i.e. 
	\begin{equation}
		\begin{split}
			\XE{\sigma}{a}{E_0}=0,& \qquad \XE{\sigma}{a}{E_B}=0,\\
			\XE{\chi}{a}{E_0}=0,& \qquad \XE{\chi}{a}{E_B}\neq 0,
		\end{split}
	\end{equation}
	where the notation  $\cdot$ represents the generators acting on the vacuum.
	Thus the only non-zero contribution to the gauge boson masses are the $\XE{\chi}{a}{E_B}$ terms. 
	Concretely, the gauge boson masses  arise from the kinetic terms
	\begin{equation}
	    \label{eq:}
	    \begin{split}
	    \mathcal{L}_{\mathrm{kin}}=&f^2\Tr[(D_{\mu}\Sigma)^{\dagger}D^{\mu}\Sigma]\\
		= & \dots+g^2f^2\Tr\left[(\XE{\chi}{a}{\Sigma})^{\dagger}(\XE{\chi}{b}{\Sigma})\right]A_{\mu}^aA^{b\, \mu},
	    \end{split}
	\end{equation}
	where
	\begin{equation}
	    \label{eq:}
	    D_{\mu}\Sigma=\partial_{\mu}\Sigma-\ii g \,A_{\mu}^a\,\XE{\tau}{a}{\Sigma}.
	\end{equation}
	Thus the gauge boson masses
	are given by
	\begin{equation}
	    \begin{split}
	    \label{eq:gaugeMass}
		(\mu^2)_{ab}=&g^2f^2\Tr\left[(\XE{\chi}{a}{\EE})^{\dagger}(\XE{\chi}{b}{\EE})\right]\\
		=&g^2f^2\sin^2\theta\Tr\left[(\XE{\chi}{a}{\EB})^{\dagger}(\XE{\chi}{b}{\EB})\right]\\	
		\equiv&g^2C(\EB)_{ab}f^2\sin^2\theta.
	    \end{split}
	\end{equation}
 The $C(\EB)_{ab}$ factor encodes the group theoretical embedding of the gauge group and, as expected, it depends only on the broken generators.

	Furthermore, we assume that the fermions acquire their masses fully from the composite condensate via four-fermion operators induced e.g. by
	extended strong dynamics, and the induced effective operators are only invariant under the gauged subgroup $K$. 
	This yields the following fermion masses  	
	\begin{equation}
	    \label{eq:}
	    m_F=\frac{y_F}{\sqrt{2}} f \sin\theta.
	\end{equation}

    %%%%%%%%%%%%%%%%%%%%%%%%%%%%%%
     \subsection{Quantum corrections}
    %%%%%%%%%%%%%%%%%%%%%%%%%%%%%%

Within the composite framework, the leading corrections  (in the EW and fermion-Yukawa couplings) to the effective potential
are given by:
\begin{equation}
    \label{eq:}
\widetilde{V}_{\mathrm{eff}}=\widetilde{V}^{\text{gauge}}_1+\widetilde{V}^{\mathrm{ferm}}_1,
\end{equation}
with
\begin{equation}
    \label{eq:}
    \begin{split}
\widetilde{V}^{\text{gauge}}_1&= \frac{3}{64\pi^2}C_K\Tr\left[ \mu^2\right]\Lambda^2, \text{ and } \\
\widetilde{V}^{\text{ferm}}_1&= -\frac{4}{64\pi^2}\sum_F C_{F}\Tr\left[ m_F^{\dagger}m_F\right]\Lambda^2,
    \end{split}
\end{equation}
where $\Lambda$ is identified with the compositeness scale $\Lambda\sim 4\pi f$, and $C_K, C_{F}$ are the form factors 
related to the gauge group $K$ and the fermion $F$, respectively.

For the SM gauge group, and top quark embedding, the contributions to the potential on the vacuum read: 
    \begin{align}
	\label{eq:VeffComp}
	    \left.\widetilde{V}^{\mathrm{gauge}}_1\right|_{\mathrm{vac}} &= \frac{3}{16}\left(3 g^2 C_g +g^{\prime 2}C_{g^{\prime}}\right)\sin^2\theta f^4,\\
	    \left.\widetilde{V}^{\mathrm{top}}_1\right|_{\mathrm{vac}}&= -\frac{3}{2}y_t^2 C_t\sin^2\theta f^4.
    \end{align} 
    The gauge contribution has a minimum at $\theta=0$, meaning that the gauge sector prefers to be unbroken in 
		 agreement with  
		Peskin~\cite{Peskin:1980gc} and Preskill~\cite{Preskill:1980mz}. 
However, the top sector prefers the minimum to be at $\theta=\pi/2$. 
This contribution was not considered in \cite{Peskin:1980gc, Preskill:1980mz} and was analysed only later in recent models of technicolor and composite-Higgs dynamics, see e.g. \cite{Galloway:2010bp,Cacciapaglia:2014uja}. Given the large top contribution, it will dominate the potential and try to align the vacuum in the direction where the electroweak symmetry is fully broken.
This means that in the composite limit, we need new sources of vacuum misalignment to achieve a pseudo-Goldstone-boson (pGB) Higgs scenario.
We will discuss such a source in the next subsection.

    %%%%%%%%%%%%%%%%%%%%%%%%%%%%%%
    \subsection{Explicit symmetry breaking }
    %%%%%%%%%%%%%%%%%%%%%%%%%%%%%%

To achieve the desired vacuum alignment, one can add an {\it ad hoc} explicit symmetry breaking operator taking the form:
\begin{equation}
	V_B = \pm\, C_B f^4 \Tr[E_0^\dagger \Sigma]=\pm C_B f^4 \cos\theta+\dots ,
\end{equation}
where $C_B$ is just a positive dimensionless  constant, and the sign depends on whether the trivial minimum is at $\theta=0$ (positive sign) or $\theta=\pi/2$ (negative sign).
Since the minimum without the explicit-breaking term is at $\theta=\pi/2$, a negative sign is needed to achieve a pGB vacuum.

The method of adding an {\it ad hoc} operator is well known in the literature 
\cite{Cacciapaglia:2014uja,Katz:2005au}.

%%%%%%%%%%%%%%%%%%%%%%%%%%%%%%%%%%%%%%%%%%%%%%%%%%%%%%%%%%%%%%%%%%%%%%%%%%%
\section{Vacuum alignment with elementary scalars}
\label{sec:elementary}
%%%%%%%%%%%%%%%%%%%%%%%%%%%%%%%%%%%%%%%%%%%%%%%%%%%%%%%%%%%%%%%%%%%%%%%%%%%

We consider an $\SO(N)$ invariant scalar potential for a scalar field $\Phi$: 
\begin{equation}
V_0  = \frac{m^2}{2} \Phi^{\dagger}\Phi + \frac{\lambda}{4!}( \Phi^{\dagger}\Phi)^2.
\end{equation}    
If $m^2$ is negative $\SO(N)$ breaks spontaneously to $\SO(N-1)$ leaving behind $N-1$ GBs. We parameterise the scalar field as
\begin{equation}
    \label{eq:Phi}
    \Phi=\left(\sigma+\ii\Pi^aX^a\right)E,
\end{equation}
where $\Pi^a$ are the GBs corresponding to the broken generators $X^a$, and $E$ gives the vacuum alignment.
The gauging of the subgroup $K$ and adding fermion-Yukawa interactions breaks the global symmetry explicitly, and thus an interesting part of the dynamics arises via quantum corrections.
For $N>4$, it is always possible to decompose the $N$ multiplet into a 4 of the $\SO(4) \cong \SU(2)_{\mathrm{L}} \times \SU(2)_{\mathrm{R}}$  and  $N-4$ singlets. The EW interactions are embedded in such a way that the  $\SO(4)$  is a subgroup of the unbroken $\SO(N-1)$. In this way  the electroweak symmetry is intact at  tree-level, and the Higgs doublet is massless.

    %%%%%%%%%%%%%%%%%%%%%%%%%%%%%%
    \subsection{The gauged subsector}
    %%%%%%%%%%%%%%%%%%%%%%%%%%%%%%
	
	Similarly as in the composite framework, Sec.~\ref{sec:composite}, we gauge $K\leq \SO(N)$ by introducing the covariant derivative
	\begin{equation}
	    \label{eq:}
	    D_{\mu}\Phi=\partial_{\mu}\Phi-\ii g \,A_{\mu}^a\,\XE{\tau}{a}{\Phi},
	\end{equation}
	and parameterise the vacuum as in Eq.~\eqref{eq:Etot}, $ E\equiv \EE=\cos\theta E_0+\sin\theta\EB$, where $E_0$ preserves $K$ and 
	$\EB$ breaks it to $K_{\mathrm{B}}$.
	Thus, when $\Phi$ acquires a vev, $\langle\Phi\rangle=f E$, the gauge bosons obtain masses 
	\begin{equation}
	    \begin{split}
		(\mu^2)_{ab}=&g^2C(\EB)_{ab}f^2\sin^2\theta.
	    \end{split}
	\end{equation}
	The $C(\EB)_{ab}$ factor encodes the group theoretical embedding of the gauge group as in the composite case.

    %%%%%%%%%%%%%%%%%%%%%%%%%%%%%%
    \subsection{Adding fermions}
    %%%%%%%%%%%%%%%%%%%%%%%%%%%%%%
	For simplicity, we consider the usual  SM-type  fermions, i.e. a Yukawa sector invariant only  under the gauge 
	group $K$ and not the full global symmetry.
	Further, we assume that the fermions obtain their masses fully via the vev of $\Phi$.	
	Then,  we can write the fermion mass-squared matrix as
	\begin{equation}
	    \label{eq:}
	    \left(m^{\dagger}m\right)_{ab}=y^2\,D(E_B)_{ab}f^2\sin^2\theta.
	\end{equation}
	Also in this case the factor $D(E_B)_{ab}$ contains information about the fermion embedding and is independent of the angle $\theta$.

    %%%%%%%%%%%%%%%%%%%%%%%%%%%%%%
    \subsection{The Coleman--Weinberg potential and the renormalisation procedure}
	\label{sec:ren}
    %%%%%%%%%%%%%%%%%%%%%%%%%%%%%%

    To illustrate the different UV structure and to relate to the composite case, let us first write down the one-loop effective potential regulated with hard cut-off, $\Lambda$:
	\begin{align}
		V_1(\bphi)& = \frac{1}{64 \pi^2}\mathrm{Str}\left[ \Lambda^4 \left( \log \Lambda^2 -\frac 1 2 \right)\right.\nonumber\\
		& + 2 \mathcal{M}^2(\bphi)\Lambda^2 \label{eq:Aurora-Helene} \\
		& \left.+  \mathcal{M}^4(\bphi) \left( \log \frac{ \mathcal{M}^2(\bphi)}{\Lambda^2}-\frac 1 2 \right)\right] +\mathrm{c.t.} \ ,\nonumber
	\end{align}	
	where $\bphi$ is the background choice, and $\mathcal{M}(\bphi)$ is the tree-level mass matrix evaluated on the given background.
	The supertrace is defined as 
\begin{equation}
\mathrm{Str}\equiv \sum_{\mathrm{scalars}}-2\sum_{\mathrm{Weyl\,\, fermions}} +3 \sum_{\mathrm{vectors}} \ .
\end{equation}

In the composite framework with a physical cut-off, the leading contributions to the effective potential, below the cut-off,  are of the same form as  the second term in Eq.~\eqref{eq:Aurora-Helene}.  In the renormalizable framework, we cancel these divergent contributions via counter terms, and the dominant contribution assumes  the form of the last term in squared brackets of Eq.~\eqref{eq:Aurora-Helene}. We choose to work  in the $\overline{\mathrm{MS}}$ scheme in which the one-loop CW potential can be written as
	\begin{equation}
	    \label{eq:Veff}
	    V_{\mathrm{eff}}=V_0+V_1^{\mathrm{scalar}}+V_1^{\mathrm{gauge}}+V_1^{\mathrm{fermion}},
	\end{equation}
	with
	\begin{align}
	    \label{eq:}
	    V_{1}^{\mathrm{scalar}}=&\frac{1}{64\pi^2}\mathrm{Tr}\left[M^4(\bphi)\left(\log\frac{M^2(\bphi)}{\mu_0^2}-\frac{3}{2}\right)
		\right],\nonumber\\
	    V_{1}^{\mathrm{gauge}}=&\frac{3}{64\pi^2}\mathrm{Tr}\left[\mu^4(\bphi)\left(\log\frac{\mu^2(\bphi)}{\mu_0^2}-\frac{5}{6}\right)
		\right],\\
	    \begin{split}
		V_{1}^{\mathrm{fermion}}=&-\frac{4}{64\pi^2}\mathrm{Tr}\left[\left(m^{\dagger}(\bphi)m(\bphi)\right)^2
		    \vphantom{\frac{3}{2}}\right.\\
		&\left.\qquad\qquad       \times \left(\log\frac{m^{\dagger}(\bphi)m(\bphi)}{\mu_0^2}-\frac{3}{2}\right)\right],
	    \end{split}\nonumber
	\end{align}
	where $M(\bphi)$, $\mu(\bphi)$, and $m(\bphi)$ are the background-dependent scalar, gauge boson, and fermion mass matrices, respectively.

	We fix the renormalisation scale such that the vev remains at the tree-level value, i.e. the one-loop tadpole contributions in the 
	$\sigma$ direction vanish, 
	\begin{equation}
	    \label{eq:renFix}
	    \left.\frac{\partial(V_1^{\mathrm{scalar}}+V_1^{\mathrm{gauge}}+V_1^{\mathrm{fermion}})}{\partial\sigma}\right|_{\sigma=f}=0.
	\end{equation}

    %%%%%%%%%%%%%%%%%%%%%%%%%%%%%%
    \subsection{The scalar and gauge contributions }
        %%%%%%%%%%%%%%%%%%%%%%%%%%%%%%

	The background-dependent masses of $\sigma$ and the gauge bosons, 
	respectively, read 
	\begin{equation}
	    \label{eq:it}
	    \begin{split}
	       &M^2_{\sigma}=\frac{\lambda}{6}(3\sigma^2-f^2),\\
		&(\mu^2)_{ab}=g^2C(\EB)_{ab}\sigma^2\sin^2\theta.
	    \end{split}
	\end{equation}    
	Thus, in the $\sigma$ direction
	\begin{align}
	    \label{eq:}
	    \begin{split}
		V_{1}^{\mathrm{scalar}}=&\frac{1}{64\pi^2}\left(\frac{\lambda}{6}(3\sigma^2-f^2)\right)^2\\
		&\times\left[\log\frac{\lambda(3\sigma^2-f^2)}{6\mu_0^2}-\frac{3}{2}\right],
	    \end{split}\nonumber\\
	    ~\\
	    \begin{split}
		V_{1}^{\mathrm{gauge}}=&\frac{3g^4}{64\pi^2}\sigma^4\sin^4\theta\\
		&\times\left[A\log\frac{g^2\sigma^2\sin^2\theta}{\mu_0^2}+B\right],\nonumber
	    \end{split}
	\end{align}
	where
	\begin{equation}
	    \label{eq:}
	    \begin{split}
		A&=\Tr[C(\EB)^2]\qquad\text{and}\\
		B&=\Tr\left[C(\EB)^2\left(\log C(\EB)-\frac{5}{6}\right)\right].
	    \end{split}
	\end{equation}	
	Fixing the renormalisation scale using  Eq.~\eqref{eq:renFix} leads to
	\begin{equation}
	    \label{eq:}
	    \begin{split}
		&\log\mu_0^2 =\frac{1}{18Ag^4\sin^4\theta+\lambda^2}\\
		&\times\left[    \vphantom{\log\frac{\lambda f^2}{3   \mathrm{e}  }}9g^4\sin^4\theta \left(2A\log(g^2f^2\sin^2\theta)\right.\right.\\
		&\quad\ \left.\left.+A+2B\right)+\lambda^2\log\frac{\lambda f^2}{3 \mathrm{e}}\right].
	    \end{split}
	\end{equation}
	Substituting this in Eq.~\eqref{eq:Veff}, we get the renormalised effective potential, $V_{\mathrm{eff}}^{\mathrm{R}}$.

    %%%%%%%%%%%%%%%%%%%%%%%%%%%%%%
    \subsection{The vacuum structure}
    %%%%%%%%%%%%%%%%%%%%%%%%%%%%%%
	The vacuum energy depends  on the angle~$\theta$. Therefore, to find the true minimum, we need 
	to further minimize with respect to $\theta$, i.e. 
	\begin{equation}
	    \label{eq:}
	\left.\frac{\partial V_{\mathrm{eff}}^{\mathrm{R}}}{\partial\theta}\right|_{\sigma=f}=0.
	\end{equation}
	This yields the following condition:

	\begin{equation}
	    \label{eq:f}
	    \begin{split}
		0=&\frac{-g^4f^4\sin^3\theta\cos\theta}{32\pi^2\left(18A\,g^4\sin^4\theta+\lambda^2\right)^2}\\
		&\times\left[\vphantom{\log\frac{3g^2\sin^2\theta}{\lambda}}972A^3g^8\sin^8\theta+90A^2g^4\lambda^2
		    \sin^4\theta\right.\\
		&\left.\ -\lambda^4\left(2A\log\frac{3g^2\sin^2\theta}{\lambda}+A+2B\right)\right].
	    \end{split}
	\end{equation}
	The potential has trivial critical points at $\theta=0$ and $\theta=\pi/2$. 
	Studying the second derivative, we find that the critical point at $\theta=0$ is  a maximum, and $\theta=\pi/2$ is a minimum.  
 
	In this work, we are 
	mainly interested in whether it is possible to determine a 
	non-trivial minimum without adding  further ingredients to the model.  
This can only exist, if
	\begin{equation}
	f(\theta)=0,
	\end{equation}
	where we defined
	
	\begin{equation}
	    \label{eq:ftheta}
	    \begin{split}
		f(\theta)\equiv &\,972A^3g^8\sin^8\theta+90A^2g^4\lambda^2\sin^4\theta\\
		    &-\lambda^4\left(2A\log\frac{3g^2\sin^2\theta}{\lambda}+A+2B\right).
	    \end{split}
	\end{equation}
	We find that $f(\theta)$ has a minimum at $\bar \theta$  for which 
	\begin{equation}
	    \label{eq:fthetaMin}
	    \sin^4\bar{\theta}=\frac{\lambda^2}{108A\,g^4},
	\end{equation}
	and at this minimum  
	\begin{equation}
	    \label{eq:}
	    f(\bar{\theta})=\frac{1}{12}\lambda^4\left[12A\log(12A)-A-24B\right].  
	\end{equation}
	In order to have a non-trivial minimum for the model, Eq. \eqref{eq:ftheta} has to be zero for some value of $\theta$.  	
	 Since $f(\theta)\rightarrow\infty$ as $\theta\rightarrow 0$ this is only possible if    $ f(\bar\theta)<0$.	    
	    Thus,
	    we obtain a condition for the desired minimum depending only on the 
	    gauge group structure:
	    \begin{equation}
		\label{eq:cond1}
		12A\log(12A)-A-24B<0.
	    \end{equation}
	    We obtain more insight if we write down $A$ and $B$ in the mass eigenbasis of the gauge bosons. To this end,  we write
	    \begin{equation}
		\label{eq:}
		\tilde{C}(E_B)_{ab}=\left(U^{-1}C(E_B)U\right)_{ab}\equiv c_a\delta_{ab}. 
	    \end{equation}
	    Then, 
	    \begin{equation}
		\label{eq:ABeig}
		\begin{split}
		    A=&\sum_ac_a^2,\quad\text{and}\quad\\
		    B=&\sum_ac_a^2\left(\log c_a-\frac{5}{6}\right),
		\end{split}
	    \end{equation}
	    and 
	    \begin{equation}
		\label{eq:cond0}
		\begin{split}
		    &12A\log (12A)-A-24B\\
		    &=\sum_a\left[12c_a^2\log\left(\sum_b 12c_b^2\right)-c_a^2\right.\\
		    &\qquad\quad\ \left.-24c_a^2\left(\log c_a-\frac{5}{6}\right)\right]\\
		    &=\sum_a\left\{12 c_a^2\log\left[ 12\left(1+\sum_{b\neq a}\frac{c_b^2}{c_a^2}\right)\right]\right.\\
		    &\qquad\quad\ \left.\vphantom{\sum_{b\neq a}\frac{c_b^2}{c_a^2}}+19c_a^2\right\}.
		\end{split}
	    \end{equation}
	    All the terms in the sum are positive, so we conclude that the condition of Eq.~\eqref{eq:cond1} can never be fulfilled, and thus 
	    there is no solution with non-vanishing vacuum alignment.
	    
	    There is still one possible caveat to be checked: If $\lambda^2>108A\,g^4$, Eq.~\eqref{eq:fthetaMin} does not have a solution, 
	    and we have to see whether $f(\pi/2)$ can be negative in this case. To do that, let $\lambda=\alpha g^2 \sqrt{108 A}$ 
	    with $\alpha>1$. Then 
	    \begin{equation}
		\label{eq:}
		\begin{split}
		    f\left(\frac{\pi}{2}\right)=& 972 A^2 g^8\left[\alpha^4\left(12 A\log(12A)-A-24B\right)\right.\\
			&\left.+A\left(12\alpha^4\log\alpha^2-11\alpha^4+10\alpha^2+1\right)\right],
		\end{split}
	    \end{equation}
	    where the first line is positive by Eq.~\eqref{eq:cond0} and the second line is always non-negative for $\alpha> 1$ (and zero 
	    at $\alpha=1$). Therefore the previous conclusion holds.

    %%%%%%%%%%%%%%%%%%%%%%%%%%%%%%
    \subsection{The fermion contribution}
\begin{figure}
	    \begin{center}
		\includegraphics[width=0.4\textwidth]{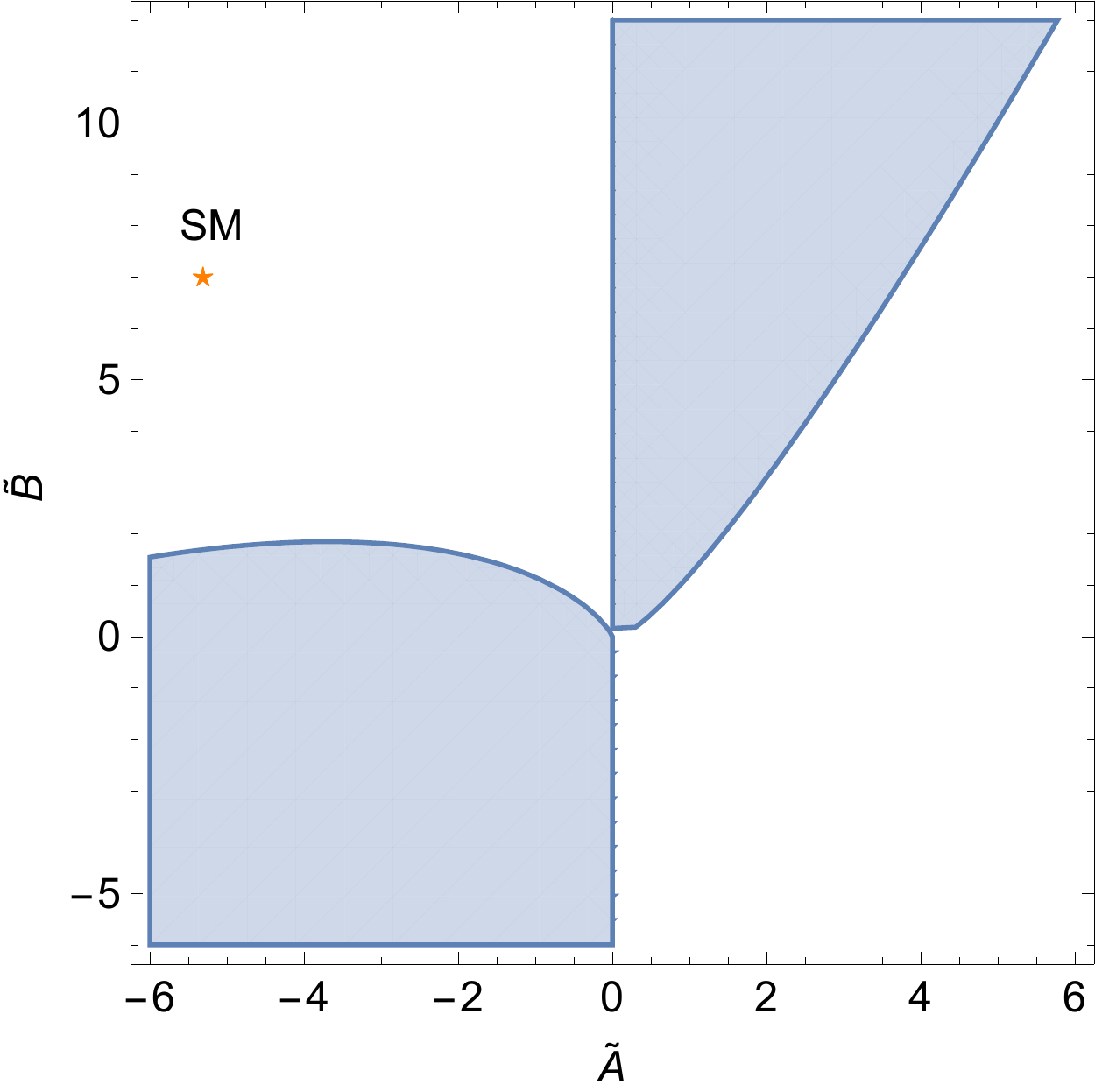}
	    \end{center}
	    \caption{The figure shows the parameter space of the theory allowing for a pseudo-Goldstone Higgs vacuum  solution (shaded region)  
		occurring for $0< \theta < \pi/2$. The plot is obtained in the minimal scenario with only the $\sigma$ field and  the one 
		loop contributions coming from the gauge and the fermion sectors. The star represents the SM embedding in this scenario,
		with the three massive gauge bosons and the top mass, see  Sec.~\ref{sec:so5} for details.
	    \label{fig:NTReg}}
	\end{figure}

    %%%%%%%%%%%%%%%%%%%%%%%%%%%%%%
    The same procedure used in the previous section can be applied, without loss of generality, to the Yukawa sector and in particular to the top quark. 
	We define, similarly to the gauge boson sector:
	\begin{equation}
	    \label{eq:}
	    \begin{split}
		A_{\mathrm{f}}&=\Tr[D(\EB)^2]\qquad\text{and}\\
		B_{\mathrm{f}}&=\Tr\left[D(\EB)^2\left(\log D(\EB)-\frac{3}{2}\right)\right],
	    \end{split}
	\end{equation}
	such that we can write the combined gauge and fermion contributions as
	\begin{equation}
	    \label{eq:ga+f}
	    \begin{split}
		&V_{1}^{\mathrm{gauge}}+V_{1}^{\mathrm{ferm}}=\\
		\frac{3g^4}{64\pi^2}&\sigma^4\sin^4\theta\left[\tilde{A}\log\frac{g^2\sigma^2\sin^2\theta}{\mu_0^2}+\tilde{B}\right],
	    \end{split}
	\end{equation}
	where we set
	\begin{equation}
	    \label{eq:tildes}
	    \begin{split}
		\tilde{A}&=A-\frac{4y^4}{3g^4}A_{\mathrm{f}}\qquad\text{and}\\
		\tilde{B}&=B-\frac{4y^4}{3g^4}B_{\mathrm{f}}-\frac{4y^4}{3g^4}A_{\mathrm{f}}\log\frac{y^2}{g^2}.
	    \end{split}
	\end{equation}
	This shows that we get the same condition for non-trivial solution as in Eq.~\eqref{eq:cond1} 
	for $\tilde{A}$ and $\tilde{B}$, if $\tilde{A}>0$. However, 
	in this case there is also the possibility to have negative $\tilde{A}$. 
	Then, following the procedure in the  previous section, we obtain 
	\begin{equation}
	    \label{eq:solfN}
	    \tilde{A}\log(-2\tilde{A})-3\tilde{A}-2\tilde{B}>0.
	\end{equation}
	The regions where these conditions are fulfilled in the $(\tilde{A},\tilde{B})$ plane are shown in Fig.~\ref{fig:NTReg}. 
	We note that in the case of dominant  fermion contributions (or neglecting gauge contributions), as in the SM-like gauge--fermion(top) sector, it is not possible to find a non-trivial solution.
	Different gauge--fermion realisations are not discussed in this work. 

In the following section, we will show, using the simplest elementary $\SO(5)\rightarrow\SO(4)$ framework,  that it is possible to 
stabilize the vacuum at a non-vanishing  value of $\theta$ by extending the scalar sector with a gauge singlet field. In this case  no explicit breaking terms are needed. Alternatively one can still add explicit symmetry breaking terms to stabilize the vacuum for a non-vanishing $\theta$.

%\comH{We don't want to do this in the general part?}
%%%%%%%%%%%%%%%%%%%%%%%%%%%%%%%
     
     %%%%%%%%%%%%%%%%%%%%%%%%%%%%%%
    \section{The $\mathrm{SO(5)\rightarrow \mathrm{SO}(4)}$ Elementary  template}
    \label{sec:so5}
    %%%%%%%%%%%%%%%%%%%%%%%%%%%%%%
	The simplest breaking pattern enabling to embed the entire Higgs doublet of the SM as GB
	is $\SO(5)\rightarrow\SO(4)$. This breaking pattern produces four GBs and it has been extensively used in
	composite Higgs models, first considered in~\cite{Agashe:2004rs,Contino:2006qr}. Here we are interested in elementary scalar degrees of freedom where  the underlying scalar potential is renormalisable. We emphasize that our treatment is different from that of~\cite{Feruglio:2016zvt}, since we consistently calculate the full one-loop potential, and determine the vacuum alignment based on these corrections.

Here and in the following, $X^a$ and $S^a$ are respectively the broken and unbroken generators normalized as $\Tr(S^a S^b) =\Tr(X^a X^b)=2\delta^{ab}$, $\Tr(S^aX^b)=0$ (the explicit expressions for the SO(5) generators can be found in Appendix \ref{app1}).
	We again identify the vacuum of the theory as a superposition of a vacuum preserving the EW group, $E_0$, and a vacuum, $E_B$, which breaks the EW group $\SU(2)_{\mathrm{L}}\times \mathrm{U}(1)_{Y}\rightarrow \mathrm{U}(1)_Q$ as $ E_\theta   = \cos\theta E_0  +\sin\theta \EB, $
	where the two vacua 
	can be explicitly written as
\begin{equation}
	E_0 =\left(0,0,0,0,1\right)^T, \quad \EB=\left(0,0,1,0,0\right)^T.
\end{equation} 
	We parameterise the $\SO(5)$ scalar multiplet similarly as in Eq.~\eqref{eq:Phi}, $\Phi=\left(\sigma+\ii\Pi^aX^a\right)E_{\theta}$.

	The masses for the $W^{\pm}$ and $Z$ bosons are:
	\begin{equation}
	    \label{eq:}
	    \begin{split}
	    \mu_{W}^2=&\frac{1}{4}g^2f^2\sin^2\theta, \\ \mu_Z^2=&\frac{1}{4}(g^2+g^{\prime\,2})f^2\sin^2\theta.
	    \end{split}
	\end{equation}
	Therefore, in order to produce the correct EW spectrum, we identify
	\begin{equation}
	    \label{eq:EWscale}
	    v_w^2=f^2\sin^2\theta.
	\end{equation}
	Finally, we couple the top quark to the left doublet of $\Phi$ such that the top quark acquires a mass
	\begin{equation}
	    \label{eq:}
	    m_t=\frac{1}{\sqrt{2}}y_t f \sin\theta.
	\end{equation}
Both the gauge bosons and fermions masses are proportional to $f \sin\theta$. 

     %%%%%%%%%%%%%%%%%%%%%%%%%%%%%%
    \subsection{Quantum corrections}
     %%%%%%%%%%%%%%%%%%%%%%%%%%%%%%

The quantum corrections are given in Eqs.~\eqref{eq:it} and~\eqref{eq:ga+f} where $\tilde A$ and $\tilde B$ now are
\begin{equation}
	\begin{split}
		\tilde A =& \frac{1}{16}\left[2+\left(1+\frac{g^{\prime\,2}}{g^2}\right)^2\right]-\frac{y^4_t}{g^4},
		\\
		\tilde B =& \frac{1}{16}\left[\left(1+\frac{g^{\prime\,2}}{g^2}\right)^2\left(\log\frac{g^2+g^{\prime\,2}}{4g^2}-\frac{5}{6}
		    \right)\right.
		\\&\left.\qquad-\frac{5}{3}-4\log 2\right]\\
		&+\frac{y^4_t}{g^4}\left(\frac{3}{2}+\log 2\right)-\frac{y^4_t}{g^4}\log \frac{y^2_t}{g^2}.
	\end{split}
\end{equation}
Making use of  the SM values for the gauge and Yukawas couplings, we obtain:
	\begin{equation}
	    \label{eq:valSM}
	   \tilde A\approx -5.2,\ \tilde B\approx 6.9.
	\end{equation}
	Note that this coincides with the star in Fig.~\ref{fig:NTReg}, showing that the EW does not break in this case, and it further illustrates the usefulness of the general results presented in the previous Section.

     %%%%%%%%%%%%%%%%%%%%%%%%%%%%%%
    \subsection{Vacuum stabilization mechanisms}
     %%%%%%%%%%%%%%%%%%%%%%%%%%%%%%

As shown above, for the SM gauge and fermion embeddings, the vacuum is never stabilized for a $\theta$ away from $0$.
In the following, we present an alternative mechanism for stabilising the vacuum in the elementary framework such that a pGB Higgs appears without adding further explicit symmetry breaking operators: We extend  the theory with a singlet scalar state. We will see that the vacuum {\it dynamically} orients itself in a direction supporting a pGB Higgs with non-vanishing value of $\theta$. 
Alternatively, we could choose to break explicitly the $\SO(5)$ symmetry to $\SO(4)$ via a minimal operator that forces the vacuum to align in the desired direction. We will briefly discuss this alternative after discussing the singlet-scalar case.

     %%%%%%%%%%%%%%%%%%%%%%%%%%%%%%
     \subsubsection{Adding a singlet }
     %%%%%%%%%%%%%%%%%%%%%%%%%%%%%% 
 
\begin{figure}
	    \begin{center}
		\includegraphics[width=0.45\textwidth]{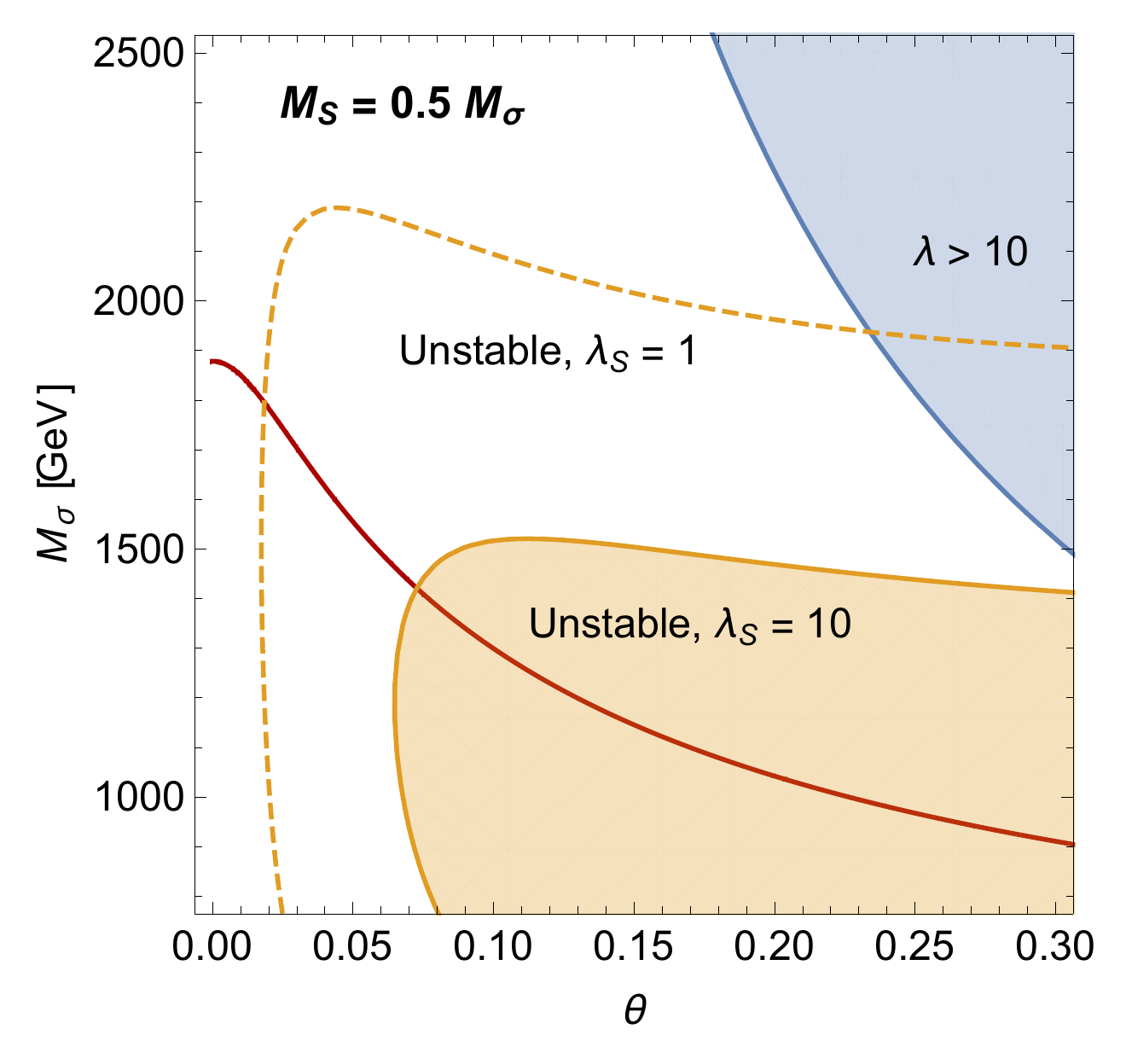}\\
		\includegraphics[width=0.45\textwidth]{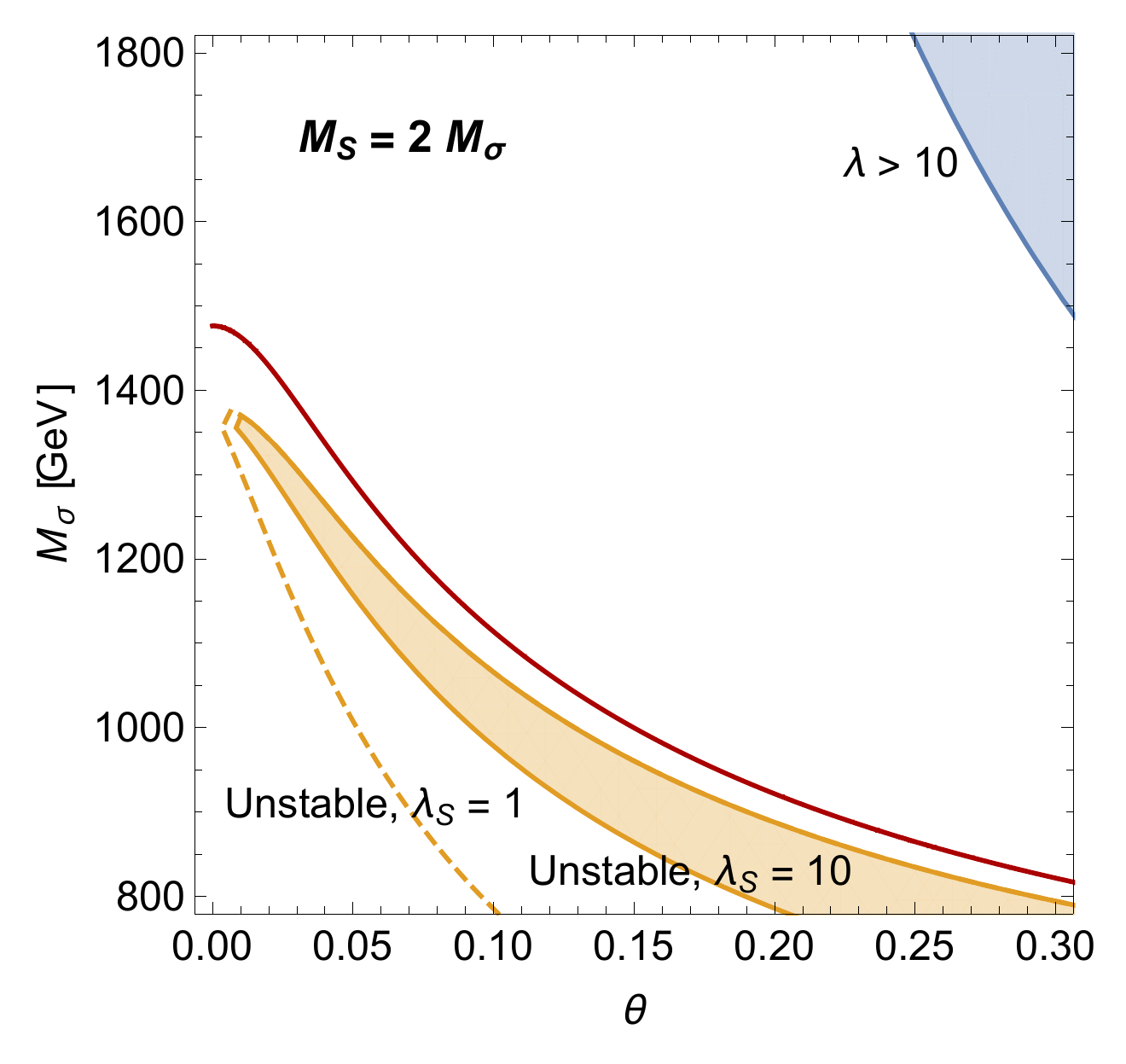}
	    \end{center}
	    \caption{The red curve gives the minimum with respect to $\theta$, and yellow (blue) area is excluded by tree-level vacuum 
	    stability (perturbativity).  We fix respectively $M_S=0.5M_{\sigma}$ ($\beta=0.5$) and $M_S=2M_{\sigma}$ ($\beta=2$)  in the upper and lower panels.}
	    \label{fig:SO5}
	\end{figure}

In the following, we show that it is possible to have a dominantly pGB Higgs with mass 125 GeV by adding a scalar singlet $S$ to the theory. For simplicity, we take this new scalar to be $Z_2$ symmetric and real. The scalar potential then reads:
\begin{equation}
	    \label{eq:}
	    \begin{split}
		V_0=&\frac{1}{2}m^2 \Phi^{\dagger}\Phi+\frac{1}{2}m_S^2S^2\\
		&+\frac{\lambda}{4!}(\Phi^\dagger \Phi)^2+\frac{\lambda_{\sigma S}}{4}(\Phi^\dagger\Phi)S^2+\frac{\lambda_S}{4!}S^4.
	    \end{split}
	\end{equation}
The stability of the potential requires:
\begin{equation} 
\label{stability}
 \lambda \geq0, ~~\lambda_S \geq0~~ {\rm and}~~ 3 \lambda_{\sigma S} +\sqrt{\lambda\,\lambda_{S}}\geq0 \ .
 \end{equation} 
Assuming for simplicity that $S$ does not acquire a vev, the background-dependent mass of the new scalar $S$ reads
\begin{equation}
	    \label{eq:}
	    \begin{split}
		M_S^2(\sigma)=&m_S^2+\frac{\lambda_{\sigma S}}{2}\sigma^2.
	    \end{split}
	\end{equation}
The singlet contributes to the one-loop effective potential with a term
    \begin{equation}
	\label{eq:}
	\begin{split}
	    V_1^S=&\frac{1}{64\pi^2}M_S^4(\sigma)\left(\log\frac{M_S^2(\sigma)}{\mu_0^2}-\frac{3}{2}\right).
	\end{split}
    \end{equation}
We then fix the renormalization scale similarly as in Eq.~\eqref{eq:renFix} and minimize again the vacuum energy with respect to 
    the alignment angle, $\theta$. Finally, we search for solutions for $\theta\in(0,\pi/2)$. To this end, it is convenient to  express the physical mass  of $S$ as $M_S=\beta M_\sigma$. Then the function $f(\theta)$  modifies to:
    
\begin{align}
    f(\theta)=& 972\tilde A^3 g^8 \sin^8\theta\nonumber\\
    & + 18\tilde A^2 g^4 \lambda  \sin^4\theta \left( 5\lambda+2 \beta^4\lambda+3 \beta^2 \lambda_S\right)\nonumber\\
    & +2\lambda^2 \left(\lambda+\beta^2\lambda_S \right)\left(2\beta^4 \lambda -\lambda -3\beta^2\lambda_S \right)\nonumber\\
    &\quad\times \left(\tilde B+\tilde A\log\frac{3g^2\sin^2\theta}{\lambda} \right)\\
    &+\tilde A\lambda^2\left[ \left( \lambda+\beta^2\lambda_S \right) \left( 8\beta^4\lambda-\lambda-9\beta^2\lambda_S \right)
      \right.\nonumber\\
    & \left.+2\beta^2\lambda_S \left( \lambda-2\beta^4\lambda+3\beta^2\lambda_S \right) \log \beta^2 \right]\nonumber,
\end{align}
where $\lambda$ is proportional to the mass squared  of $\sigma$, $\lambda=6\frac{M_\sigma^2}{f^2}$.
Finding a vacuum requires $f(\theta)$ to vanish under the constraints that the Higgs mass is properly reproduced.     This leaves us with two free parameters besides $\theta$ (in addition to $\lambda_S$ which does not affect the vacuum 
	alignment and the Higgs mass constraint) which we choose to be the tree-level masses of $\sigma$ and $S$, $M_{\sigma}$ and $M_{S}$, respectively. 
	We further fix the mass of $S$ to $0.5 M_{\sigma}$ or $2 M_{\sigma}$ as benchmark values. The parameter space is further 
	constrained by the requirement of (tree-level) vacuum stability and perturbativity of the quartic couplings ($\lambda \ll 4\pi$). 
	We show the results in Fig.~\ref{fig:SO5}.
 We observe that along the continuous red line in the $M_{\sigma} - \theta$ space, there is a ground state. 
 It is useful to note that  using Eqs.~\eqref{eq:it} and~\eqref{eq:EWscale}, one roughly expects $M_{\sigma} \propto \sqrt{\lambda}/\sin{\theta}$ on the ground state.
 In addition, we require overall stability of the potential expressed in Eq.~\eqref{stability}. It is clear that for small or negative $\lambda_{\sigma S}$,  one needs to have a sufficiently large $\lambda_S$, as it is clear from the top panel of Fig.~\ref{fig:SO5}. We are guaranteed that the overall solution occurs for perturbative values of $\lambda$ since large values of this coupling (the light blue region in the top corner of the plots) do not admit  a ground state. The same analysis is repeated for $M_S=2M_{\sigma}$ in the bottom panel of Fig.~\ref{fig:SO5}.
 
  The numerical result shows that the introduction of a singlet,  dynamically misaligns the vacuum to a value of $\theta \neq 0$. 
     
Last, it is instructive to investigate the decoupling limit for the singlet $S$, i.e. $M_S\gg v$. 
In this limit we obtain:
\begin{equation} 
\label{eq:lambda-theta}
\begin{split}
	\sin^2\theta = & \lambda_{\sigma S}\frac{v_{w}^2  \left( 3 A + 2 B + 2 A \log \dfrac{g^2 v_{w}^2}{M_S^2}\right)}{4 M_S^2 \left( 2 A + B + A \log  \dfrac{g^2 v_{w}^2}{M_S^2}\right) }\\
	&\quad + {\cal O} \left(\left({f/M_S}\right)^{4}\right).
\end{split}
\end{equation}
As expected in the exact decoupled limit the EW symmetry remains intact. For finite values of the $S$ mass the portal coupling is directly responsible for a non-vanishing vacuum value of $\theta$. In this sense, this mechanism has a dynamical origin.

     %%%%%%%%%%%%%%%%%%%%%%%%%%%%%%
    \subsubsection{Explicit symmetry breaking}
     %%%%%%%%%%%%%%%%%%%%%%%%%%%%%%

Adding an {\it ad hoc} explicit symmetry-breaking term can be done similarly as in the composite case by adding the operator of the form:
\begin{equation}
	V_B = \pm\, C_B f^3 E_0^\dagger \Phi,
\end{equation}
where $C_B$ is again a positive dimensionless  constant.
Contrary to the composite case, we are now interested in moving  the minimum away from zero, so the {\it ad hoc} operator has to be positive. In the $\sigma$ background, 
$V_B$ reads
\begin{equation}
	V_B= C_B f^3 \sigma \cos\theta.
\end{equation}

\begin{figure}
	    \begin{center}
		\includegraphics[width=0.45\textwidth]{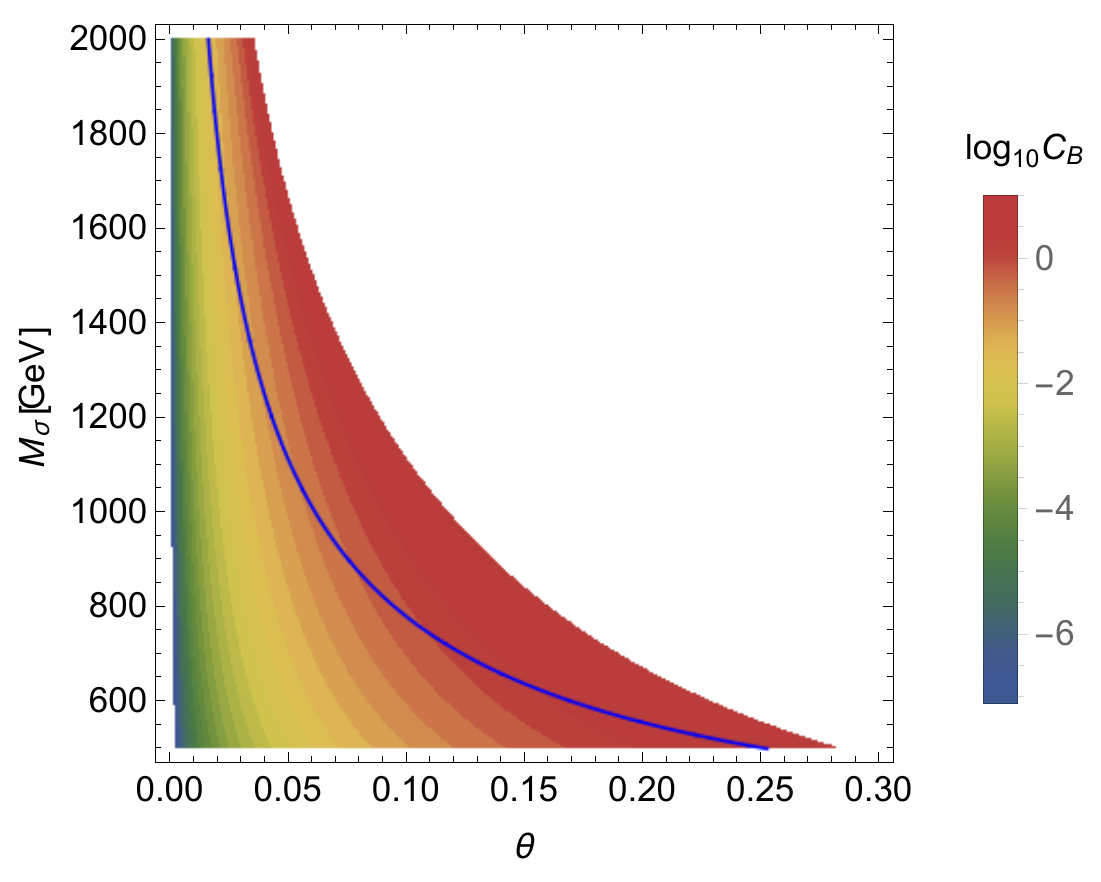}
	    \end{center}
	    \caption{The plot shows the stationary points of the potential with respect to $\theta$ (blue solid curve) when the explicit symmetry-breaking term, $V_B$, is introduced.  The coloured region corresponds to positive $C_B$  and the colour gradient indicates the value of $\log_{10}C_B$. }
	    \label{fig:adhoc}
	\end{figure}

Imposing the correct mass of the Higgs and minimizing the full potential as previously, we find that there  are now solutions for $\theta$ different from zero. In Fig.~\ref{fig:adhoc} we plot the solution of the minimum of the potential with respect to $\theta$ as function of $\theta$ and of the mass of the heavy state $\sigma$, $M_\sigma$. The coloured region of the plot corresponds to positive $C_B$ and the colour gradient indicates the  value of $\log_{10}C_B$. From the plot we can see that for $M_\sigma> 500$ GeV, the value of  $\theta$ is smaller than 0.25. We observe also that 
$M_\sigma$ and  $\theta$ are inversely proportional.

    %%%%%%%%%%%%%%%%%%%%%%%%%%%%%%
    \subsection{Minimal elementary Goldstone Higgs and dark matter: $\SU(4)\cong\SO(6)\rightarrow \Sp(4) \cong \SO(5)$}
    %%%%%%%%%%%%%%%%%%%%%%%%%%%%%%
	This scenario was studied in detail in~\cite{Alanne:2014kea,Gertov:2015xma}; here we just summarise the main results. The breaking 
	pattern is achieved with a scalar multiplet, $M$, transforming under $6_{\mathrm{A}}$ of $\SU(4)$. However, 
	since $6_{\mathrm{A}}$ is real, and 
	the breaking pattern is	locally isomorphic to $\SO(6)\rightarrow\SO(5)$, we know that non-trivial vacuum alignment cannot be 
	achieved with minimal scalar degrees of freedom. However, we can start from the six-dimensional complex representation 
	of $\mathrm{U}(4)$ and break the $\mathrm{U}(4)$-invariant potential to $\SU(4)$ by introducing Pfaffian terms. 
	This gives 12 scalar decrees of freedom and the spectrum consists of two $\Sp(4)$ singlets, $\sigma$ and its pseudoscalar partner $\Theta$, and two $\Sp(4)$ quintuplets, the pions $\Pi^i$ and their (massive) scalar partners $\tilde{\Pi}^i$.

	The left and right generators of $\SU(2)_{\mathrm{L}}\times \SU(2)_{\mathrm{R}}$ embedded in the $\SU(4)$ are identified with

	\begin{equation}
	    \label{eq:gensCust}
	    T^i_{\mathrm{L}}=\frac{1}{2}\left(\begin{array}{cc}\sigma_i & 0 \\ 0 & 0\end{array}\right),\qquad
	    T^i_{\mathrm{R}}=\frac{1}{2}\left(\begin{array}{cc} 0 & 0 \\ 0 & -\sigma_i^{T}\end{array}\right),
	\end{equation}
	where $\sigma_i$ are the Pauli matrices. The generator of the hypercharge is then identified with the third generator 
	of the $\SU(2)_{\mathrm{R}}$ group, $T_Y=T^3_{\mathrm{R}}$. Further, the vacuum that leaves EW intact is given by\footnote{As 
	discussed in~\cite{Galloway:2010bp}, there are actually two inequivalent vacua of this kind, but with either choice, the physical 
	properties of the pGBs are the same.}
	\beq
	E_0 = \left( \begin{array}{cc}
	\ii \sigma_2 & 0 \\
	0 & - \ii \sigma_2
	\end{array} \right)\,,
	\eeq
	while the alignment that breaks the EW symmetry to $\mathrm{U}(1)_Q$ is given by
	\beq
	E_B  =\left( \begin{array}{cc}
	0 & 1 \\
	-1 & 0
	\end{array} \right) \, .
	\eeq
	Following the discussion in the previous sections, we define the vacuum of the theory  following Eq. \eqref{eq:Etot}
	such that $E_{\theta}^{\dagger} E_{\theta}^{\vphantom{\dagger}} = 1$.
	The EW group is gauged by introducing the covariant derivative 
	\begin{equation}
	    \label{eq:covM}
	    D_{\mu}M=\partial_{\mu}M-\ii\left(G_{\mu}M+MG_{\mu}^{\mathrm{T}}\right),
	\end{equation}
	where
	\begin{equation}
	    \label{eq:gaugeG}
	    G_{\mu}=gW^i_{\mu}T_{\mathrm{L}}^i+g'B_{\mu}T^3_{\mathrm{R}}\, ,
	\end{equation}
	and the generators $T^i_{\mathrm{L}}$ and $T^3_{\mathrm{R}}$ are given by Eq.~\eqref{eq:gensCust}. 
	When $\sigma$ acquires a vacuum expectation value, $f$, the weak gauge bosons acquire masses
	\begin{equation}
	    \label{eq:WBosMasses}
	    \begin{split}
		m_W^2=&\frac{1}{4}g^2f^2\sin^2\theta, \quad\text{and}\\
		m_Z^2=&\frac{1}{4}(g^2+g'^2)f^2\sin^2\theta.
	    \end{split}
	\end{equation}  
	Further, we couple the SM fermions, in particular the top quark, to the EW doublet within $M$. The Yukawa term is then given 
	by~\cite{Galloway:2010bp}
	\begin{equation}
	    \label{eq:topYuk}
	    \mathcal{L}_{\mathrm{Yuk}}=y_t(Qt^{c})^{\dagger}_{\alpha}\mathrm{Tr}[P_{\alpha}M]+\mathrm{h.c.},
	\end{equation}
	where $P_{1,2}$ pick the components of the $\SU(2)_{\mathrm{L}}$ doublet.
	The top quark then acquires the following mass:
	\begin{equation}
	    \label{eq:mtop}
	    m_t=\frac{y_t}{\sqrt{2}}f\sin\theta.
	\end{equation}
	
\begin{figure}
	    \begin{center}
		\includegraphics[width=0.4\textwidth]{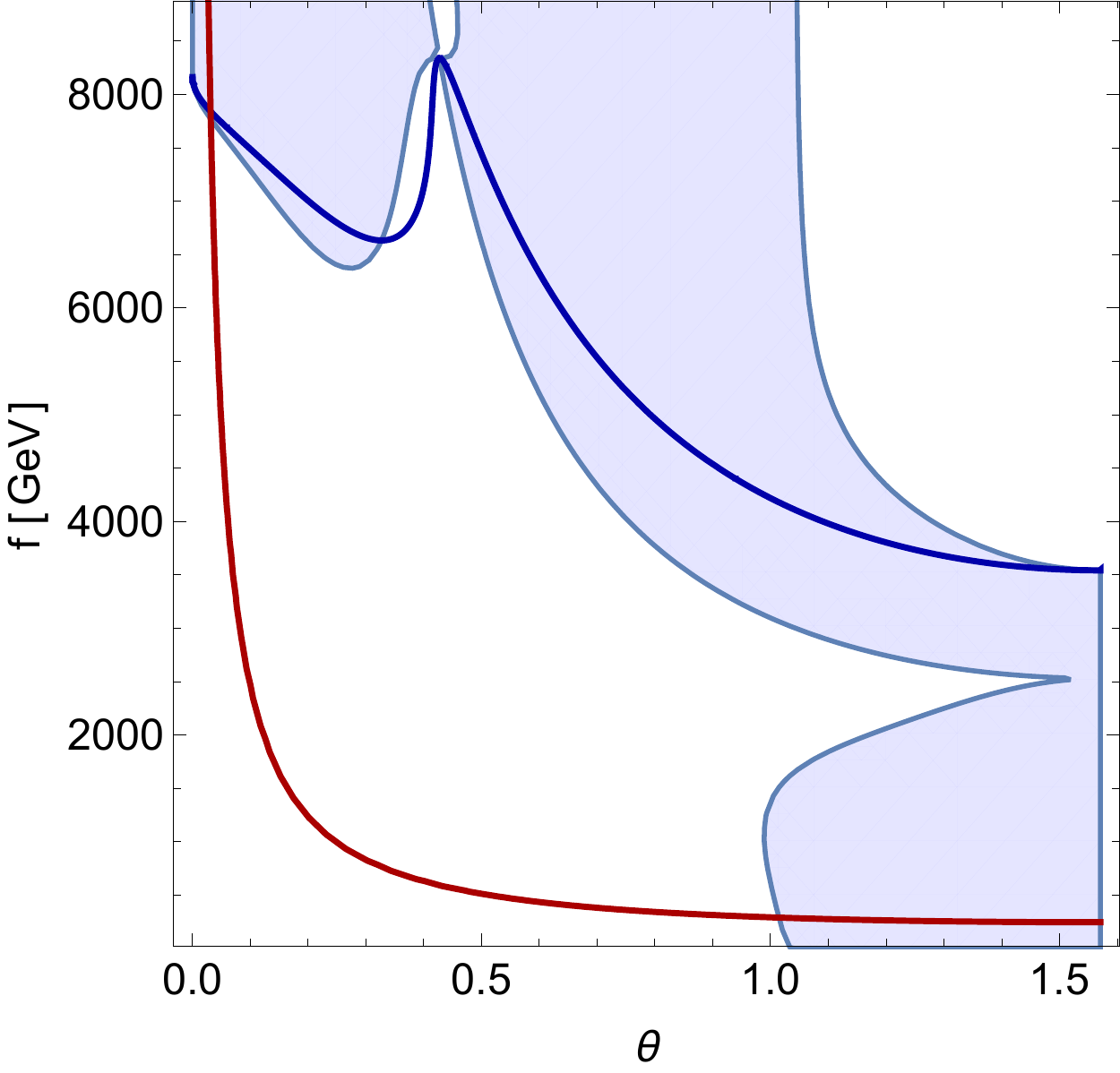}
	    \end{center}
	    \caption{The blue contour shows the stationary points with respect to $\theta$, and on the blue regions the second derivative
	    with respect to $\theta$ is positive, so the blue contour on a blue region represents a minimum. The red contour represents the 
	    points giving the correct EW gauge boson and top quark masses.
	    For details, see~\cite{Alanne:2014kea}.}
	    \label{fig:EGH}
	\end{figure}

	In the simplest case, where all the massive scalars have equal tree-level masses, the vacuum alignment depends on one 
	effective quartic coupling, $\tilde{\lambda}$, and we find phenomenologically viable solutions with $\theta\ll 1$, 
	i.e. $f$ being in  the multi-TeV range~\cite{Alanne:2014kea,Gertov:2015xma}. 
	Furthermore, the DM candidate can account for the full observed DM abundance, and be consistent with the experimental constraints
	in the region $m_{\mathrm{DM}}>125$~GeV. 
	A benchmark scenario for the vacuum alingment with $\tilde{\lambda}=0.1$ is shown in 
	Fig.~\ref{fig:EGH}.
	
	In practice the dynamics, for this example, is similar to the case in which we added a singlet state. In other words the potential of the theory is sufficiently rich to allow for a ground state with a non-vanishing $\theta$.

%%%%%%%%%%%%%%%%%%%%%%%%%%%%%%%%%%%%%%%%%%%%%%%%%%%%%%%%%%%%%%%%%%%%%%%%%%%
\section{Conclusions}
\label{sec:conclusions}
%%%%%%%%%%%%%%%%%%%%%%%%%%%%%%%%%%%%%%%%%%%%%%%%%%%%%%%%%%%%%%%%%%%%%%%%%%%
We have investigated the vacuum alignment problem in renormalizable extensions of the SM that can feature a pGB Higgs. We have shown that the structure of the calculable radiative corrections differs from the composite-Goldstone-Higgs paradigm yielding different ways in which the vacuum stabilization mechanisms work. We provided sufficiently general conditions showing, for example,  that renormalizable theories with a single massive scalar singlet cannot radiatively stabilize the vacuum. However when generalizing the theory by including at least one more massive scalar state, the vacuum can be made stable  without the aid of other mechanisms such as the introduction of new global-symmetry-breaking operators often invoked in composite extensions. We then  applied our results to phenomenologically relevant examples.

\section*{Acknowledgments}
 The CP3-Origins centre is partially funded by the Danish National Research Foundation, grant number DNRF90.  
 TA acknowledges partial funding from a Villum foundation grant.

\appendix
\section{SO(5)/SO(4) generators}\label{app1}
	We first identify the $\SU(2)_{\mathrm{L}}\times\SU(2)_{\mathrm{R}}$ subgroup of $\SO(5)$ by fixing the left and right
	generators as 
	\begin{equation}
	    \label{eq:}
	    \begin{split}
		\left(T_{\mathrm{L,R}}\right)_{ij}^a=-\frac{\ii}{2}&\left[\frac{1}{2}\epsilon^{abc}\left(\delta^b_i\delta^c_j-
		    \delta^b_j\delta^c_i\right)\right.\\
		    &\left. \ \pm\left(\delta_i^a\delta_j^4-\delta_j^a\delta_i^4\right)\right],
	    \end{split}
	\end{equation}
	where $a=1,2,3$ and $i,j=1,\dots , 5$.
	The generator 
	$T_{\mathrm{R}}^3$ is then identified as the generator of hypercharge.
We list here  the SO(5) generators adopted in this work. 
	\begin{equation}
	    \label{eq:}
	    \begin{split}
		X^{1}_{ij}=&-\ii\left[\sin\theta \left(\delta_i^1\delta_j^3-\delta_i^3\delta_j^1\right)\right.\\
		    &\left.\qquad+\cos\theta\left(\delta_i^1\delta_j^5-\delta_i^5\delta_j^1\right)\right],\\
		X^{2}_{ij}=&-\ii\left[\sin\theta \left(\delta_i^2\delta_j^3-\delta_i^3\delta_j^2\right)\right.\\
		    &\left.\qquad+\cos\theta\left(\delta_i^2\delta_j^5-\delta_i^5\delta_j^2\right)\right],\\
		X^{3}_{ij}=&-\ii\left[-\sin\theta \left(\delta_i^3\delta_j^4-\delta_i^4\delta_j^3\right)\right.\\
		    &\left.\qquad+\cos\theta\left(\delta_i^4\delta_j^5-\delta_i^5\delta_j^4\right)\right],\\
		X^{4}_{ij}=&-\ii\left(\delta_i^3\delta_j^5-\delta_i^5\delta_j^3\right).
	    \end{split}	
	\end{equation}
The generators are normalised such that $\Tr[X^a_{\theta}X^b_{\theta}]=2\delta^{ab}$.

\end{document}